\documentclass{article}
\usepackage{aas_macros}
\usepackage{microtype}
\usepackage{graphicx}
\usepackage{subfigure}
\usepackage{booktabs} 

\usepackage{hyperref}

\usepackage[accepted]{icml2025}

\usepackage{amsmath}
\usepackage{amssymb}
\usepackage{mathtools}
\usepackage{amsthm}

\usepackage[capitalize,noabbrev]{cleveref}

\theoremstyle{plain}

\theoremstyle{definition}

\theoremstyle{remark}

\usepackage[textsize=tiny]{todonotes}

\icmltitlerunning{VADER: Inferring Planet and Disk Properties}

\begin{document}

\twocolumn[
\icmltitle{VADER: A Variational Autoencoder to Infer Planetary Masses and Gas-Dust Disk Properties Around Young Stars}




\begin{icmlauthorlist}
\icmlauthor{Sayed Shafaat Mahmud}{col}
\icmlauthor{Sayantan Auddy}{fi}
\icmlauthor{Neal Turner}{jpl}
\icmlauthor{Jeffrey S. Bary}{col}
\end{icmlauthorlist}

\icmlaffiliation{col}{Colgate University, 13 Oak Drive, Hamilton, NY 13346, USA}
\icmlaffiliation{fi}{Fields Institute, 222 College St Toronto, ON M5T 3J1, Canada}
\icmlaffiliation{jpl}{Jet Propulsion Laboratory, California Institute of Technology, 4800 Oak Grove Drive, Pasadena, CA 91109, USA}

\icmlcorrespondingauthor{Sayed Shafaat Mahmud}{smahmud@colgate.edu}

\icmlkeywords{Variational Autoencoders, Protoplanetary Disks, Planet Formation, Bayesian Inference, Astrophysics, Deep Learning, Exoplanets}

\vskip 0.3in
]



\printAffiliationsAndNotice{\icmlEqualContribution} 

\begin{abstract}
We present \textbf{VADER} (Variational Autoencoder for Disks Embedded with Rings), for inferring both planet mass and global disk properties from high-resolution ALMA dust continuum images of protoplanetary disks (PPDs). VADER, a probabilistic deep learning model, enables uncertainty-aware inference of planet masses, $\alpha$-viscosity, dust-to-gas ratio, Stokes number, flaring index, and the number of planets directly from protoplanetary disk images. VADER is trained on over 100{,}000 synthetic images of PPDs generated from \texttt{FARGO3D} simulations post-processed with \texttt{RADMC3D}. Our trained model predicts physical planet and disk parameters with $R^2 > 0.9$ from dust continuum images of PPDs. Applied to 23 real disks, VADER's mass estimates are consistent with literature values and reveal latent correlations that reflect known disk physics. Our results establish VAE-based generative models as robust tools for probabilistic astrophysical inference, with direct applications to interpreting protoplanetary disk substructures in the era of large interferometric surveys.
\end{abstract}

\section{Introduction}
\label{sec: Intro}

Understanding how planets form and evolve requires interpreting the rich substructures seen in high-resolution images of protoplanetary disks. With the advent of ALMA, disks are now routinely resolved into rings, gaps, and spirals—features that potentially trace the dynamical influence of forming planets \citep{andrews2018scaling, isella2018signatures,huang2020large, fedele2018alma, bae2022structured, cieza2021ophiuchus}. However, connecting these observed morphologies to physical quantities such as planet mass, disk viscosity, or dust properties remains a formidable inverse problem \citep{haworth2016grand, rosotti2023empirical, dominik2024bouncing, jiang2024grain}. Traditional approaches rely on running large suites of hydrodynamic simulations and comparing their outputs to observations—a process that is computationally intensive and difficult to scale \citep{toci2019long, veronesi2019multiwavelength, veronesi2021dynamical, teague2018kinematical}. Moreover, many disk features are inherently degenerate: multiple combinations of parameters can produce nearly identical images \citep{kanagawa2016mass, dipierro2015dust, lodato2019newborn, zhang2018disk, long2020dual}.

Machine learning has recently emerged as a promising alternative, offering faster inference by learning mappings from images to physical parameters \citep{aud20,auddy2022using, ribas2020modeling}. Most of these methods, however, are based on conventional models like multi-layer perceptrons \cite{aud20} or convolutional neural networks \citep{auddy2021dpnnet, zhang2022pgnets, telkamp2022machine}, which yield deterministic estimates and fail to capture the prediction uncertainties.

In this work, we present \textbf{VADER}, a Variational Autoencoder framework that addresses these gaps by modeling the full posterior distribution over planetary and disk parameters from resolved ALMA-like disk images. Trained on synthetic data derived from \texttt{FARGO3D} and \texttt{RADMC3D}, our model encodes disk morphology into a smooth latent space and infers both local and global physical quantities—including masses of up to three embedded planets and disk properties such as $\alpha$-viscosity and dust-to-gas ratio. Unlike prior models, VADER produces uncertainty-aware predictions and generalizes effectively to real observations. This work lays the foundation for probabilistic inference in planet and disk properties.

\section{Methods}\label{sec: Methods}

\subsection{Synthetic Dataset Generation}

We train our model on a dataset of over 100{,}000 synthetic images of protoplanetary disks. These images are generated using the \texttt{FARGO3D} hydrodynamic code and post-processed with the \texttt{RADMC3D} radiative transfer package. The dataset emulates ALMA Band 6 dust continuum observations at 1300\,$\mu$m. The simulations include up to three embedded planets and span a wide range of physical parameters: planet masses (8\,$M_\oplus$ to 3\,$M_\mathrm{J}$), $\alpha$-viscosity (10$^{-4}$ to 5$\times$10$^{-2}$), dust-to-gas ratio (0.01 to 0.05), Stokes number (10$^{-4}$ to 1.57), and flaring index (0.01 to 0.25), where $M_\oplus$ is the mass of Earth and $M_\mathrm{J}$ is the mass of Jupiter. Each simulation produces $\sim$150 augmented images with randomized inclinations, position angles, and translations, resulting in a diverse training set of axisymmetric disk morphologies.

\begin{figure}[ht]
\vskip 0.2in
\begin{center}
\centerline{\includegraphics[width=\columnwidth]{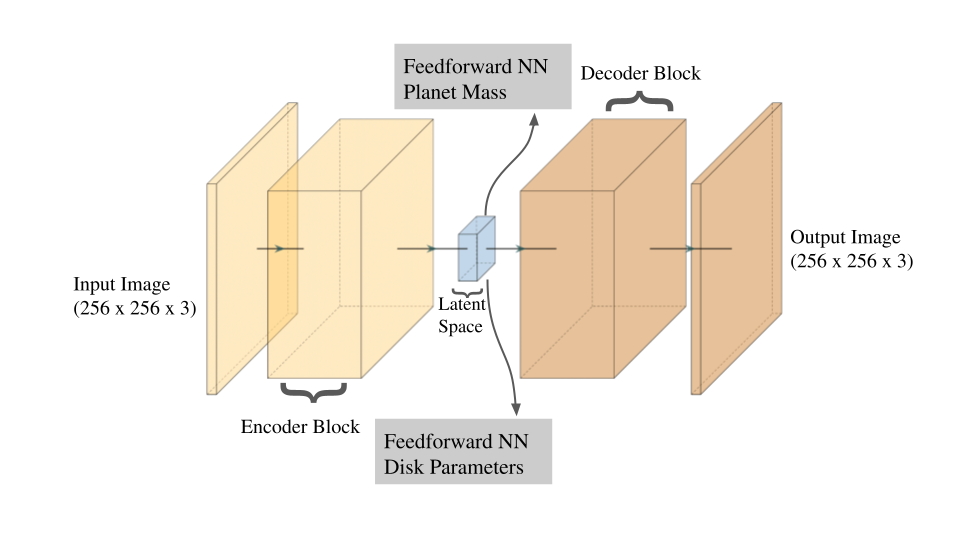}}
\caption{Simplified schematic of the variational autoencoder (VAE) architecture used in this work. The encoder compresses the input dust continuum image of a protoplanetary disk (\(256 \times 256 \times 3\)) into a 32-dimensional latent space. Two feedforward neural networks independently process the latent vector to infer planetary masses and disk parameters, respectively. The decoder reconstructs the input image from the latent representation.}
\label{fig: vae_architecture}
\end{center}
\vskip -0.2in
\end{figure}

\subsection{Model Architecture}

Our framework consists of a Variational Autoencoder (VAE) paired with two feedforward neural networks (FNNs) for parameter inference. The VAE compresses a 256$\times$256 RGB disk image into a 32-dimensional latent vector via a 3-layer convolutional encoder. The decoder reconstructs the original image using transposed convolutions. From the latent vectors, one FNN predicts the masses of up to three planets (regression), while the second predicts disk properties such as $\alpha$-viscosity, dust-to-gas ratio, Stokes number, flaring index, and number of planets (mapped to ordinal labels, treated as regression targets). Both FNNs consist of 7 hidden layers with batch normalization and ReLU activations.

\subsection{Training Objective}
The VAE model is trained by minimizing the reconstruction and regularization loss terms
\begin{equation}
    \mathcal{L}_{\text{VAE}} = \mathbb{E}_{q_\phi(z|x)}[\log p_\theta(x|z)] - D_{\text{KL}}(q_\phi(z|x) \, || \, p(z)),
\end{equation}
where $q_\phi(z|x)$ is the encoder, $p_\theta(x|z)$ is the decoder, and $p(z) = \mathcal{N}(0, I)$ is the prior over the latent space. The encoder maps the high-dimensional input image \( x \) into a latent space \( z \). The goal of the decoder is to reconstruct the original input \( x \) from \( z \). We use mean squared error (MSE) for the reconstruction loss and KL divergence for regularization. 
After training the VAE, the two FNNs are trained independently. Each FNN receives the 32-dimensional latent vectors from the trained VAE as input and predicts planet mass or disk properties by minimizing the MSE loss. 

\subsection{Optimization Details}

All networks are trained using the Adam optimizer with a learning rate of $10^{-3}$ and batch size of 32. We train for 150 epochs and select the best checkpoint based on validation set performance. During evaluation, we sample multiple latent vectors per image to obtain predictive uncertainties for each parameter.

\section{Results}

We evaluate VADER on both synthetic test data and real ALMA observations, assessing reconstruction quality, parameter inference accuracy, and uncertainty quantification.

\subsection{Synthetic Performance}

\begin{figure}[ht]
\vskip 0.2in
\begin{center}
\centerline{\includegraphics[width=1.06\columnwidth]{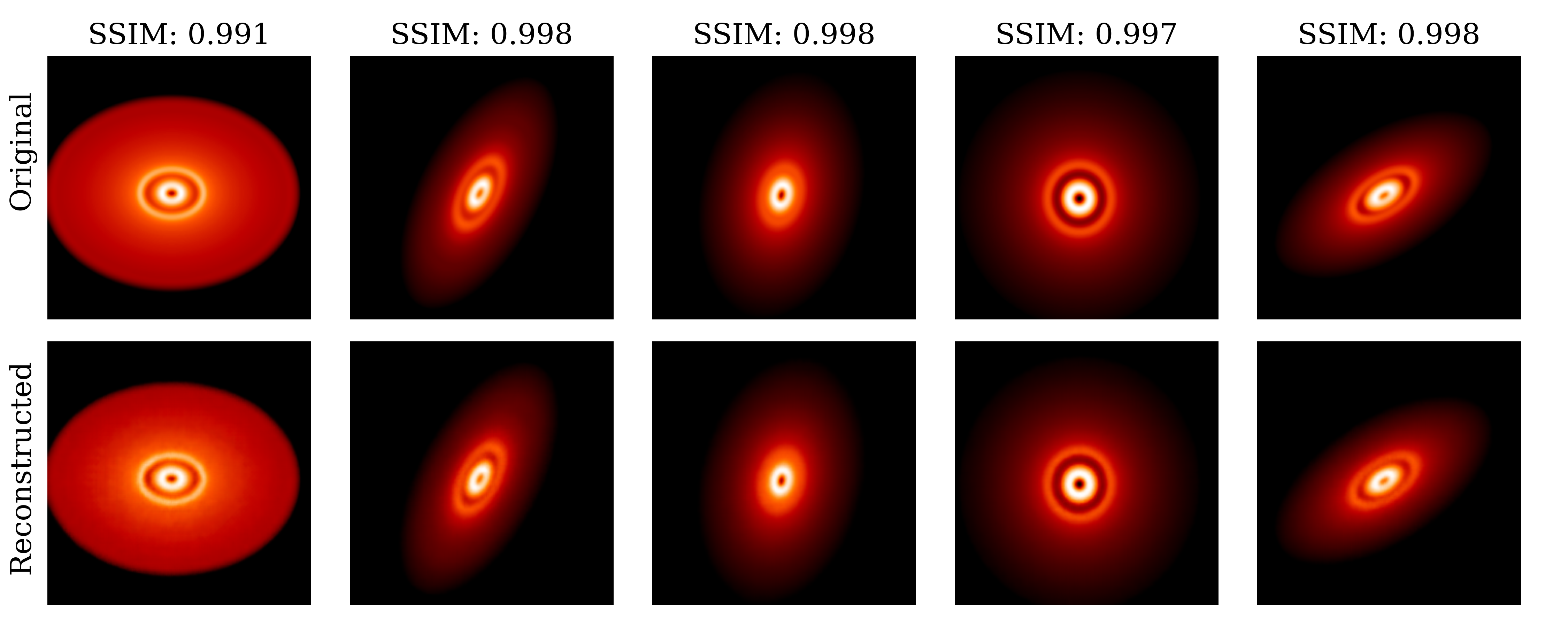}}
\caption{Comparison of original (top row) and reconstructed (bottom row) sample images from the test set using our VAE model. Each image was randomly selected, and the SSIM between the original and reconstructed images exceeds $0.99$.}
\label{fig: SSIM_reconstruction}
\end{center}
\vskip -0.2in
\end{figure}

Figure~\ref{fig: SSIM_reconstruction} shows examples of input and reconstructed disk images. VADER achieves a mean Structural Similarity Index (SSIM) exceeding $0.99$ across the test set, indicating near-perfect morphological recovery of complex structures including concentric rings, asymmetries, and gaps.

\begin{figure}[ht]
\vskip 0.2in
\begin{center}
\centerline{\includegraphics[width=1.06\columnwidth]{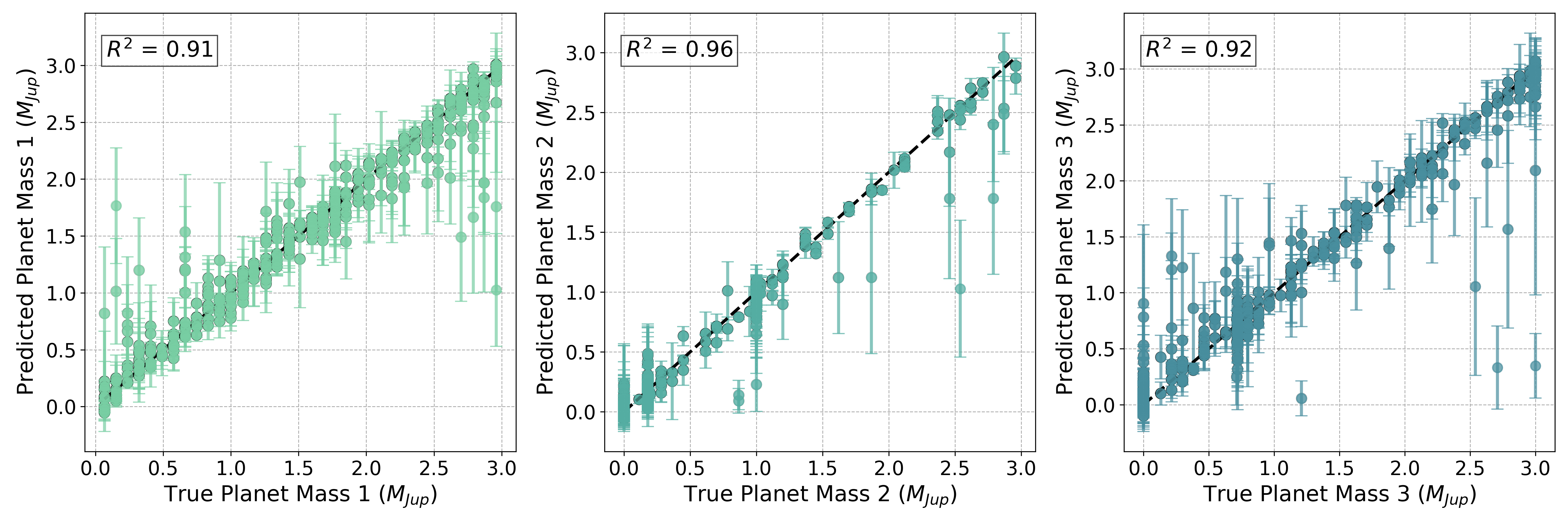}}
\caption{Actual vs. predicted values for the mass of each of the three possible embedded planets, with error bars indicating uncertainty in the predictions. For all cases, the coefficient of determination ($R^2$) exceeds 0.9, demonstrating strong predictive performance.}
\label{fig: mass_pred}
\end{center}
\vskip -0.2in
\end{figure}

To assess parameter inference, we compare predicted and ground-truth values for planetary masses and disk parameters. For planet masses, VADER achieves $R^2 > 0.9$ (Figure~\ref{fig: mass_pred}), with uncertainties growing proportionally as the predicted masses diverge from the true values. For uncertainty quantification, we sample 500 latent points from the distribution formed by the 32 dimensional latent space and pass each through the FNNs which after 500 iterations give us the posterior distribution for a given parameter.

\begin{figure}[ht]
\vskip 0.2in
\begin{center}
\centerline{\includegraphics[width=1.06\columnwidth]{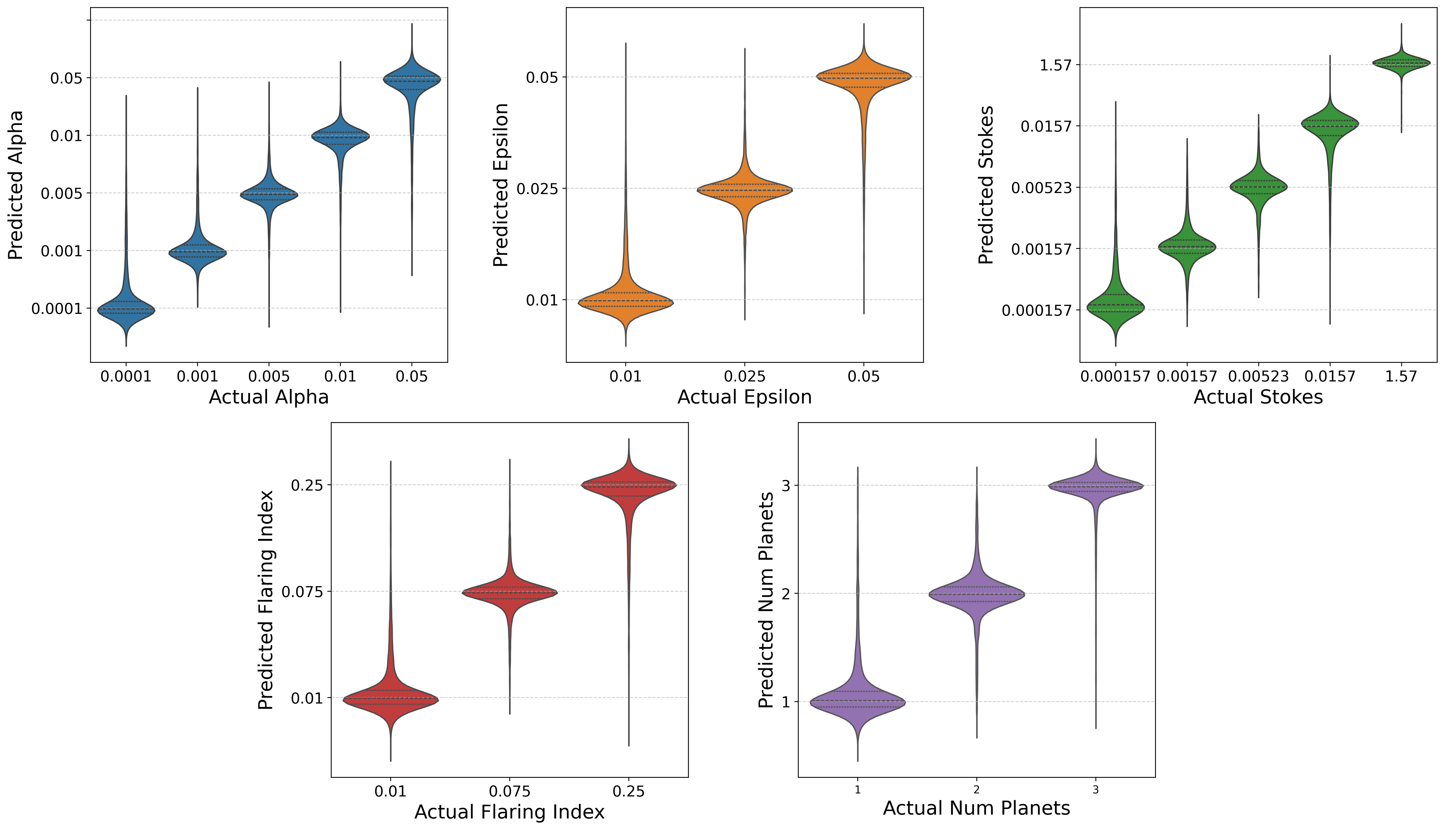}}
\caption{Violin plots comparing predicted and actual parameter values for protoplanetary disks across five disk parameters in the test data: $\alpha$-viscosity (top left), dust-to-gas ratio $\epsilon$ (top center), Stokes number (top right), flaring index (bottom left), and number of embedded planets (bottom right). For each parameter, the violin shows the distribution of corresponding predictions made by our model. The interior dashed lines within each violin denote the 25th, 50th (median), and 75th percentiles of the predictive distribution, illustrating both the central tendency and spread.}
\label{fig: disk_params}
\end{center}
\vskip -0.2in
\end{figure}

Disk parameters are drawn from a discrete set but treated via ordinal regression. Violin plots (Figure~\ref{fig: disk_params}) show the distributions of predicted values for each true category (e.g., $\alpha$-viscosity, dust-to-gas ratio, Stokes number). Predictions are tightly concentrated around correct classes with minimal category overlap, capturing both ordinal structure and physical scaling. This suggests that VADER can reliably recover global disk properties despite inherent morphological degeneracies.

\subsection{Application to Observational Data}

\begin{figure*}[ht!]
    \centering
    \includegraphics[width=0.9\textwidth]{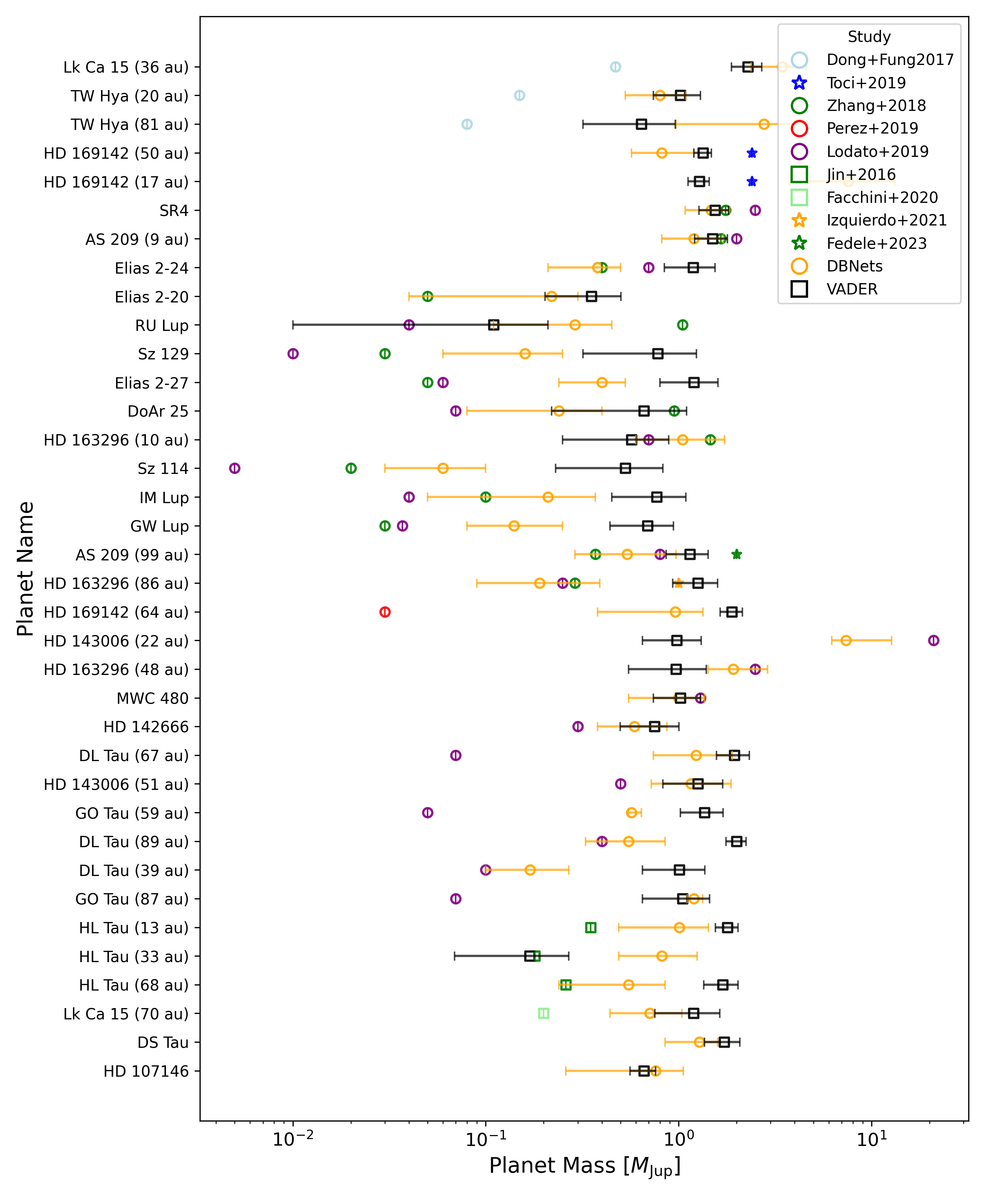}
    \caption{Comparison of predicted planet masses from our model with estimates from previous studies across different protoplanetary systems \citep{dong2017mass, toci2019long, zhang2018disk, perez2019dust, lodato2019newborn, jin2016modeling, facchini2020annular, izquierdo2022new, fedele2018alma, ruzza2024dbnets}. The error bars indicate the uncertainty in mass predictions, with our VAE-based predictions shown in black squares. A key validation of our model is that many of its predicted masses are within 1$\sigma$ of the values reported in prior studies, demonstrating strong agreement with independent analyses. Additionally, the predictions from DBNets (orange circles) are shown for reference, further supporting the consistency of our results. }
    \label{fig:vae_planet_mass_comparison}
\end{figure*}

We apply VADER to 23 protoplanetary disks observed with ALMA and benchmark against literature values. Figure~\ref{fig:vae_planet_mass_comparison} compares our predicted planet masses (black squares) with prior estimates from DBNets~\citep{ruzza2024dbnets} and hydrodynamic modeling studies~\citep{zhang2018disk, dong2017mass}. Our predictions agree within $1\sigma$ of published values for the majority of systems. This supports the generalization capability of the model beyond synthetic training distributions. We perform disk parameter predictions on two of the most studied protoplanetary disks: HL Tau and HD 163296

For HL Tau, VADER predicts three embedded planets with masses $1.81\pm0.23\,M_{\mathrm{Jup}}$, $0.18\pm0.11\,M_{\mathrm{Jup}}$, and $1.72\pm0.35\,M_{\mathrm{Jup}}$, all consistent with ALMA gap locations and prior inferences~\citep{jin2016modeling}. Disk parameters such as $\alpha \sim 10^{-3} - 5 \times 10^{-3}$, $\epsilon \sim 0.025$ and Stokes Number $\sim 10^{-3} - 10^{-2}$ align with expectations from disk evolution models~\citep{pinte2015dust, wu2018physical} and aligns well with planet formation scenarios. 

In HD 163296, VADER identifies three planets of masses $0.571 \pm 0.3~M_{\text{Jup}}$, $0.95 \pm 0.44~M_{\text{Jup}}$, and $1.25 \pm 0.33~M_{\text{Jup}}$, in agreement with predictions from \cite{zhang2018disk}. It infers $\alpha \sim 10^{-3}$ and Stokes numbers $\sim 10^{-3} - 5 \times 10^{-3}$, matching ranges suggested by CO kinematic analysis and dust trapping studies~\citep{teague2018kinematical, rodenkirch2021modeling}. Dust-to-gas ratios inferred by our model fall between $1\%$ and $2.5\%$, consistent with multi-wavelength modeling~\citep{isella2016ringed}.

\subsection{Scientific Implications}

VADER uncovers a population of sub-Jovian to Jovian planets at wide separations (10--100~AU), consistent with expectations from disk–planet interaction theory. The ability to yield probabilistic parameters from images, positions VADER as a useful tool to study and categorize newly forming disk-planet systems in the era of high-resolution ALMA imaging.


\section*{Impact Statement}

This paper presents work whose goal is to advance the field of machine learning and its application to astrophysical inference. We introduce VADER, the first variational autoencoder framework capable of full uncertainty-aware parameter inference on protoplanetary disk observations. Our model can infer disk and planetary parameters in seconds, significantly accelerating analysis compared to traditional simulation-based pipelines. By enabling fast, probabilistic interpretation of high-resolution observations, VADER opens new avenues for population-level studies of planet formation.

In future applications, this framework can be extended to support multimodal astronomy by integrating heterogeneous observational inputs—including dust continuum maps, infrared emission, and CO gas velocity fields—into a unified latent space. Such a capability would enhance the fidelity of astrophysical inference by combining complementary constraints from diverse data sources.

Our method is designed for scientific use and does not involve human-related data or downstream decision-making systems. We do not foresee ethical risks or negative societal impacts arising from this work. Nonetheless, we encourage responsible deployment and transparent communication when using machine learning models for scientific discovery.
\section*{Acknowledgments}
This work was carried out in part at the Jet Propulsion Laboratory, California Institute of Technology, under contract with NASA.
\bibliographystyle{icml2025}
\bibliography{example_paper}

\begin{thebibliography}{35}
\providecommand{\natexlab}[1]{#1}
\providecommand{\url}[1]{\texttt{#1}}
\expandafter\ifx\csname urlstyle\endcsname\relax
  \providecommand{\doi}[1]{doi: #1}\else
  \providecommand{\doi}{doi: \begingroup \urlstyle{rm}\Url}\fi

\bibitem[Andrews et~al.(2018)Andrews, Terrell, Tripathi, Ansdell, Williams, and Wilner]{andrews2018scaling}
Andrews, S.~M., Terrell, M., Tripathi, A., Ansdell, M., Williams, J.~P., and Wilner, D.~J.
\newblock Scaling relations associated with millimeter continuum sizes in protoplanetary disks.
\newblock \emph{The Astrophysical Journal}, 865\penalty0 (2):\penalty0 157, 2018.

\bibitem[{Auddy} \& {Lin}(2020){Auddy} and {Lin}]{aud20}
{Auddy}, S. and {Lin}, M.-K.
\newblock {A Machine Learning Model to Infer Planet Masses from Gaps Observed in Protoplanetary Disks}.
\newblock \emph{\apj}, 900\penalty0 (1):\penalty0 62, September 2020.
\newblock \doi{10.3847/1538-4357/aba95d}.

\bibitem[Auddy et~al.(2021)Auddy, Dey, Lin, and Hall]{auddy2021dpnnet}
Auddy, S., Dey, R., Lin, M.-K., and Hall, C.
\newblock Dpnnet-2.0. i. finding hidden planets from simulated images of protoplanetary disk gaps.
\newblock \emph{The Astrophysical Journal}, 920\penalty0 (1):\penalty0 3, 2021.

\bibitem[Auddy et~al.(2022)Auddy, Dey, Lin, Carrera, and Simon]{auddy2022using}
Auddy, S., Dey, R., Lin, M.-K., Carrera, D., and Simon, J.~B.
\newblock Using bayesian deep learning to infer planet mass from gaps in protoplanetary disks.
\newblock \emph{The Astrophysical Journal}, 936\penalty0 (1):\penalty0 93, 2022.

\bibitem[Bae et~al.(2022)Bae, Isella, Zhu, Martin, Okuzumi, and Suriano]{bae2022structured}
Bae, J., Isella, A., Zhu, Z., Martin, R., Okuzumi, S., and Suriano, S.
\newblock Structured distributions of gas and solids in protoplanetary disks.
\newblock \emph{arXiv preprint arXiv:2210.13314}, 2022.

\bibitem[Cieza et~al.(2021)Cieza, Gonz{\'a}lez-Ruilova, Hales, Pinilla, Ru{\'\i}z-Rodr{\'\i}guez, Zurlo, Casassus, P{\'e}rez, C{\'a}novas, Arce-Tord, et~al.]{cieza2021ophiuchus}
Cieza, L.~A., Gonz{\'a}lez-Ruilova, C., Hales, A.~S., Pinilla, P., Ru{\'\i}z-Rodr{\'\i}guez, D., Zurlo, A., Casassus, S., P{\'e}rez, S., C{\'a}novas, H., Arce-Tord, C., et~al.
\newblock The ophiuchus disc survey employing alma (odisea)--iii. the evolution of substructures in massive discs at 3--5 au resolution.
\newblock \emph{Monthly Notices of the Royal Astronomical Society}, 501\penalty0 (2):\penalty0 2934--2953, 2021.

\bibitem[Dipierro et~al.(2015)Dipierro, Pinilla, Lodato, and Testi]{dipierro2015dust}
Dipierro, G., Pinilla, P., Lodato, G., and Testi, L.
\newblock Dust trapping by spiral arms in gravitationally unstable protostellar discs.
\newblock \emph{Monthly Notices of the Royal Astronomical Society}, 451\penalty0 (1):\penalty0 974--986, 2015.

\bibitem[Dominik \& Dullemond(2024)Dominik and Dullemond]{dominik2024bouncing}
Dominik, C. and Dullemond, C.
\newblock The bouncing barrier revisited: Impact on key planet formation processes and observational signatures.
\newblock \emph{Astronomy \& Astrophysics}, 682:\penalty0 A144, 2024.

\bibitem[Dong \& Fung(2017)Dong and Fung]{dong2017mass}
Dong, R. and Fung, J.
\newblock What is the mass of a gap-opening planet?
\newblock \emph{The Astrophysical Journal}, 835\penalty0 (2):\penalty0 146, 2017.

\bibitem[Facchini et~al.(2020)Facchini, Benisty, Bae, Loomis, Perez, Ansdell, Mayama, Pinilla, Teague, Isella, et~al.]{facchini2020annular}
Facchini, S., Benisty, M., Bae, J., Loomis, R., Perez, L., Ansdell, M., Mayama, S., Pinilla, P., Teague, R., Isella, A., et~al.
\newblock Annular substructures in the transition disks around lkca 15 and j1610.
\newblock \emph{Astronomy \& Astrophysics}, 639:\penalty0 A121, 2020.

\bibitem[Fedele et~al.(2018)Fedele, Tazzari, Booth, Testi, Clarke, Pascucci, Kospal, Semenov, Bruderer, Henning, et~al.]{fedele2018alma}
Fedele, D., Tazzari, M., Booth, R., Testi, L., Clarke, C., Pascucci, I., Kospal, A., Semenov, D., Bruderer, S., Henning, T., et~al.
\newblock Alma continuum observations of the protoplanetary disk as 209-evidence of multiple gaps opened by a single planet.
\newblock \emph{Astronomy \& Astrophysics}, 610:\penalty0 A24, 2018.

\bibitem[Haworth et~al.(2016)Haworth, Ilee, Forgan, Facchini, Price, Boneberg, Booth, Clarke, Gonzalez, Hutchison, et~al.]{haworth2016grand}
Haworth, T.~J., Ilee, J.~D., Forgan, D.~H., Facchini, S., Price, D.~J., Boneberg, D.~M., Booth, R.~A., Clarke, C.~J., Gonzalez, J.-F., Hutchison, M.~A., et~al.
\newblock Grand challenges in protoplanetary disc modelling.
\newblock \emph{Publications of the Astronomical society of Australia}, 33:\penalty0 e053, 2016.

\bibitem[Huang et~al.(2020)Huang, Andrews, {\"O}berg, Ansdell, Benisty, Carpenter, Isella, P{\'e}rez, Ricci, Williams, et~al.]{huang2020large}
Huang, J., Andrews, S.~M., {\"O}berg, K.~I., Ansdell, M., Benisty, M., Carpenter, J.~M., Isella, A., P{\'e}rez, L.~M., Ricci, L., Williams, J.~P., et~al.
\newblock Large-scale co spiral arms and complex kinematics associated with the t tauri star ru lup.
\newblock \emph{The Astrophysical Journal}, 898\penalty0 (2):\penalty0 140, 2020.

\bibitem[Isella \& Turner(2018)Isella and Turner]{isella2018signatures}
Isella, A. and Turner, N.~J.
\newblock Signatures of young planets in the continuum emission from protostellar disks.
\newblock \emph{The Astrophysical Journal}, 860\penalty0 (1):\penalty0 27, 2018.

\bibitem[Isella et~al.(2016)Isella, Guidi, Testi, Liu, Li, Li, Weaver, Boehler, Carperter, De~Gregorio-Monsalvo, et~al.]{isella2016ringed}
Isella, A., Guidi, G., Testi, L., Liu, S., Li, H., Li, S., Weaver, E., Boehler, Y., Carperter, J.~M., De~Gregorio-Monsalvo, I., et~al.
\newblock Ringed structures of the hd 163296 protoplanetary disk revealed by alma.
\newblock \emph{Physical Review Letters}, 117\penalty0 (25):\penalty0 251101, 2016.

\bibitem[Izquierdo et~al.(2022)Izquierdo, Facchini, Rosotti, van Dishoeck, and Testi]{izquierdo2022new}
Izquierdo, A.~F., Facchini, S., Rosotti, G.~P., van Dishoeck, E.~F., and Testi, L.
\newblock A new planet candidate detected in a dust gap of the disk around hd 163296 through localized kinematic signatures: An observational validation of the discminer.
\newblock \emph{The Astrophysical Journal}, 928\penalty0 (1):\penalty0 2, 2022.

\bibitem[Jiang et~al.(2024)Jiang, Mac{\'\i}as, Guerra-Alvarado, and Carrasco-Gonz{\'a}lez]{jiang2024grain}
Jiang, H., Mac{\'\i}as, E., Guerra-Alvarado, O.~M., and Carrasco-Gonz{\'a}lez, C.
\newblock Grain-size measurements in protoplanetary disks indicate fragile pebbles and low turbulence.
\newblock \emph{Astronomy \& Astrophysics}, 682:\penalty0 A32, 2024.

\bibitem[Jin et~al.(2016)Jin, Li, Isella, Li, and Ji]{jin2016modeling}
Jin, S., Li, S., Isella, A., Li, H., and Ji, J.
\newblock Modeling dust emission of hl tau disk based on planet--disk interactions.
\newblock \emph{The Astrophysical Journal}, 818\penalty0 (1):\penalty0 76, 2016.

\bibitem[Kanagawa et~al.(2016)Kanagawa, Muto, Tanaka, Tanigawa, Takeuchi, Tsukagoshi, and Momose]{kanagawa2016mass}
Kanagawa, K.~D., Muto, T., Tanaka, H., Tanigawa, T., Takeuchi, T., Tsukagoshi, T., and Momose, M.
\newblock Mass constraint for a planet in a protoplanetary disk from the gap width.
\newblock \emph{Publications of the Astronomical Society of Japan}, 68\penalty0 (3):\penalty0 43, 2016.

\bibitem[Lodato et~al.(2019)Lodato, Dipierro, Ragusa, Long, Herczeg, Pascucci, Pinilla, Manara, Tazzari, Liu, et~al.]{lodato2019newborn}
Lodato, G., Dipierro, G., Ragusa, E., Long, F., Herczeg, G.~J., Pascucci, I., Pinilla, P., Manara, C.~F., Tazzari, M., Liu, Y., et~al.
\newblock The newborn planet population emerging from ring-like structures in discs.
\newblock \emph{Monthly Notices of the Royal Astronomical Society}, 486\penalty0 (1):\penalty0 453--461, 2019.

\bibitem[Long et~al.(2020)Long, Pinilla, Herczeg, Andrews, Harsono, Johnstone, Ragusa, Pascucci, Wilner, Hendler, et~al.]{long2020dual}
Long, F., Pinilla, P., Herczeg, G.~J., Andrews, S.~M., Harsono, D., Johnstone, D., Ragusa, E., Pascucci, I., Wilner, D.~J., Hendler, N., et~al.
\newblock Dual-wavelength alma observations of dust rings in protoplanetary disks.
\newblock \emph{The Astrophysical Journal}, 898\penalty0 (1):\penalty0 36, 2020.

\bibitem[P{\'e}rez et~al.(2019)P{\'e}rez, Casassus, Baruteau, Dong, Hales, and Cieza]{perez2019dust}
P{\'e}rez, S., Casassus, S., Baruteau, C., Dong, R., Hales, A., and Cieza, L.
\newblock Dust unveils the formation of a mini-neptune planet in a protoplanetary ring.
\newblock \emph{The Astronomical Journal}, 158\penalty0 (1):\penalty0 15, 2019.

\bibitem[Pinte et~al.(2015)Pinte, Dent, M{\'e}nard, Hales, Hill, Cortes, and de~Gregorio-Monsalvo]{pinte2015dust}
Pinte, C., Dent, W.~R., M{\'e}nard, F., Hales, A., Hill, T., Cortes, P., and de~Gregorio-Monsalvo, I.
\newblock Dust and gas in the disk of hl tauri: surface density, dust settling, and dust-to-gas ratio.
\newblock \emph{The Astrophysical Journal}, 816\penalty0 (1):\penalty0 25, 2015.

\bibitem[Ribas et~al.(2020)Ribas, Espaillat, Mac{\'\i}as, and Sarro]{ribas2020modeling}
Ribas, {\'A}., Espaillat, C.~C., Mac{\'\i}as, E., and Sarro, L.~M.
\newblock Modeling protoplanetary disk seds with artificial neural networks-revisiting the viscous disk model and updated disk masses.
\newblock \emph{Astronomy \& Astrophysics}, 642:\penalty0 A171, 2020.

\bibitem[Rodenkirch et~al.(2021)Rodenkirch, Rometsch, Dullemond, Weber, and Kley]{rodenkirch2021modeling}
Rodenkirch, P.~J., Rometsch, T., Dullemond, C.~P., Weber, P., and Kley, W.
\newblock Modeling the nonaxisymmetric structure in the hd 163296 disk with planet-disk interaction.
\newblock \emph{Astronomy \& Astrophysics}, 647:\penalty0 A174, 2021.

\bibitem[Rosotti(2023)]{rosotti2023empirical}
Rosotti, G.~P.
\newblock Empirical constraints on turbulence in proto-planetary discs.
\newblock \emph{New Astronomy Reviews}, 96:\penalty0 101674, 2023.

\bibitem[Ruzza et~al.(2024)Ruzza, Lodato, and Rosotti]{ruzza2024dbnets}
Ruzza, A., Lodato, G., and Rosotti, G.~P.
\newblock Dbnets: A publicly available deep learning tool to measure the masses of young planets in dusty protoplanetary discs.
\newblock \emph{Astronomy \& Astrophysics}, 685:\penalty0 A65, 2024.

\bibitem[Teague et~al.(2018)Teague, Bae, Bergin, Birnstiel, and Foreman-Mackey]{teague2018kinematical}
Teague, R., Bae, J., Bergin, E.~A., Birnstiel, T., and Foreman-Mackey, D.
\newblock A kinematical detection of two embedded jupiter-mass planets in hd 163296.
\newblock \emph{The Astrophysical Journal Letters}, 860\penalty0 (1):\penalty0 L12, 2018.

\bibitem[Telkamp et~al.(2022)Telkamp, Mart{\'\i}nez-Palomera, Duch{\^e}ne, Ashimbekova, Wolfe, Angelo, and Pinte]{telkamp2022machine}
Telkamp, Z., Mart{\'\i}nez-Palomera, J., Duch{\^e}ne, G., Ashimbekova, A., Wolfe, E., Angelo, I., and Pinte, C.
\newblock A machine learning framework to predict images of edge-on protoplanetary disks.
\newblock \emph{The Astrophysical Journal}, 939\penalty0 (2):\penalty0 73, 2022.

\bibitem[Toci et~al.(2019)Toci, Lodato, Fedele, Testi, and Pinte]{toci2019long}
Toci, C., Lodato, G., Fedele, D., Testi, L., and Pinte, C.
\newblock Long-lived dust rings around hd 169142.
\newblock \emph{The Astrophysical Journal Letters}, 888\penalty0 (1):\penalty0 L4, 2019.

\bibitem[Veronesi et~al.(2019)Veronesi, Lodato, Dipierro, Ragusa, Hall, and Price]{veronesi2019multiwavelength}
Veronesi, B., Lodato, G., Dipierro, G., Ragusa, E., Hall, C., and Price, D.
\newblock Multiwavelength observations of protoplanetary discs as a proxy for the gas disc mass.
\newblock \emph{Monthly Notices of the Royal Astronomical Society}, 489\penalty0 (3):\penalty0 3758--3768, 2019.

\bibitem[Veronesi et~al.(2021)Veronesi, Paneque-Carreno, Lodato, Testi, P{\'e}rez, Bertin, and Hall]{veronesi2021dynamical}
Veronesi, B., Paneque-Carreno, T., Lodato, G., Testi, L., P{\'e}rez, L.~M., Bertin, G., and Hall, C.
\newblock A dynamical measurement of the disk mass in elias 2--27.
\newblock \emph{The Astrophysical Journal Letters}, 914\penalty0 (2):\penalty0 L27, 2021.

\bibitem[Wu et~al.(2018)Wu, Hirano, Takakuwa, Yen, and Aso]{wu2018physical}
Wu, C.-J., Hirano, N., Takakuwa, S., Yen, H.-W., and Aso, Y.
\newblock Physical and chemical conditions of the protostellar envelope and the protoplanetary disk in hl tau.
\newblock \emph{The Astrophysical Journal}, 869\penalty0 (1):\penalty0 59, 2018.

\bibitem[Zhang et~al.(2018)Zhang, Zhu, Huang, Guzm{\'a}n, Andrews, Birnstiel, Dullemond, Carpenter, Isella, P{\'e}rez, et~al.]{zhang2018disk}
Zhang, S., Zhu, Z., Huang, J., Guzm{\'a}n, V.~V., Andrews, S.~M., Birnstiel, T., Dullemond, C.~P., Carpenter, J.~M., Isella, A., P{\'e}rez, L.~M., et~al.
\newblock The disk substructures at high angular resolution project (dsharp). vii. the planet--disk interactions interpretation.
\newblock \emph{The Astrophysical journal letters}, 869\penalty0 (2):\penalty0 L47, 2018.

\bibitem[Zhang et~al.(2022)Zhang, Zhu, and Kang]{zhang2022pgnets}
Zhang, S., Zhu, Z., and Kang, M.
\newblock Pgnets: planet mass prediction using convolutional neural networks for radio continuum observations of protoplanetary discs.
\newblock \emph{Monthly Notices of the Royal Astronomical Society}, 510\penalty0 (3):\penalty0 4473--4484, 2022.

\end{thebibliography}




\end{document}